\documentclass[showpacs,twocolumn,amssymb,pre,aps]{revtex4}
\usepackage{graphicx,epsfig}
\usepackage{color}
\usepackage{comment}

\newcommand{\dC}{$^{\circ}$C}

\begin{document}
\title{Thermotropic Nematic Order Upon Nano-Capillary Filling}
\author{Patrick Huber$^{1,2,3}$ }
\author{Mark Busch$^{1}$ }
\author{Sylwia~Ca{\l}us$^{4}$ }
\author{Andriy~V.~Kityk$^{4}$ }
\affiliation{$^1$ Materials Physics and Technology, Hamburg University of Technology, D-21073 Hamburg-Harburg, Germany\\
 $^2$ Department of Biomaterials, Max-Planck Institute of Colloids and Interfaces, D-14424 Potsdam, Germany\\
 $^3$ Experimental Physics, Saarland University, Saarbr\"ucken, Germany\\
 $^4$ Faculty of Electrical Engineering, Czestochowa University of Technology, P-42200 Czestochowa, Poland}

\date{\today}

\begin{abstract}
Optical birefringence and light absorption measurements reveal four regimes for the thermotropic behavior of a nematogen liquid (7CB) upon sequential filling of parallel-aligned capillaries of 12~nm diameter in a monolithic, mesoporous silica membrane. No molecular reorientation is observed for the first adsorbed monolayer. In the film-condensed state (up to 1~nm thickness) a weak, continuous paranematic-to-nematic (P-N) transition is found, which is shifted by 10~K below the discontinuous bulk transition at T$_{\rm IN}=$305~K. The capillary-condensed state exhibits a more pronounced, albeit still continuous P-N reordering, located 4~K below T$_{\rm IN}$. \textcolor{black}{This shift vanishes abruptly on complete filling of the capillaries. It could originate in competing anchoring conditions at the free inner surfaces and at the pore walls or result from the 10~MPa tensile pressure release associated with the disappearance of concave menisci in the confined liquid upon complete filling. The study documents that the thermo-optical properties of nanoporous systems (or single nano-capillaries) can be tailored over a surprisingly wide range simply by variation of the the filling fraction with liquid crystals}.
\end{abstract}


\maketitle

\section{Introduction}
Liquid crystals spatially confined on the nanometer scale exhibit structural and thermodynamical properties which differ markedly from their bulk counterparts. Both the collective orientational (isotropic-to-nematic) and translational (smectic) transitions have turned out to be significantly by finite size, quenched disorder and interfacial (solid-liquid) interactions introduced by the confining solid walls or the pore topology \cite{Sheng1976,Crawford1996,Bellini1992,Iannacchione1993,Guegan2007,Qian1998,Kutnjak2003,Kutnjak2004, Grigoriadis2011, Araki2011, Ruths2012}.
 
For example, the first-order discontinuous isotropic-to-nematic transition of rod-like liquid crystals condensed in pores a few nanometers across goes to a critical point, above which, strictly speaking, it disappears. Then the transition is rendered continuous with a paranematic high-temperature phase, where even at highest temperatures the pore walls (anchoring fields) induce a residual collective orientational, paranematic order and no isotropic liquid is observable \cite{Iannacchione1993, Kutnjak2003, Kutnjak2004, Kityk2008}.

This sensitivity of liquid crystalline phase transitions stimulated a plethora of experimental studies. Typically they were performed upon complete filling of the confining geometry \cite{Crawford1996, Iannacchione1993, Kutnjak2003, Kutnjak2004, Kityk2008,Grigoriadis2011} and thus revealed integral quantities on the phase behavior. Phenomenological considerations \cite{Sheng1976} and computer simulations on liquid crystals in thin film and nanopore confinement \cite{Gruhn1997, Gruhn1998, Binder2008, Ji2009a, Ji2009b,Pizzirusso2012} indicate, however, pronounced heterogeneities, in particular interface-induced molecular layering and orientational immobility in the pore wall proximity. Despite recent experimental advancements in optical techniques directly probing orientational order parameter profiles in the proximity of planar, solid walls \cite{Barna2008, DeLuca2008, Lee2009}, achieving the spatial resolution necessary to rigorously explore such inhomogeneities in nanometer-sized capillaries remains still experimentally extremely demanding. 

Here we present an experimental scheme which allows us to investigate the thermotropic orientational order of a classic, rod-like nematogen (7CB) as a function of gradual filling of parallel-aligned silica capillaries in a monolithic membrane. We find a gradual evolution of the thermotropic orientational order behavior and can infer four distinct phase transition behaviors depending on the filling fraction, and thus spatial arrangement of the liquid crystalline material in the capillaries. Moreover, we find hints of tensile pressure effects on the orientational behavior of the nano-confined liquid crystal. 

\section{Experimental}
In our experiments we employ a monolithic silica membrane of 320 $\mu$m thickness. It is traversed by an array of parallel-aligned, non-interconnected capillaries. The membrane is prepared by thermal oxidation of a free-standing mesoporous silicon membrane \cite{Gruener2008a, Kumar2008} at 810\dC~for 12~h. The mean capillary diameter $D$ and porosity $P$, as determined by recording a volumetric nitrogen sorption isotherm at $T=$77~K, are $12.0$ $\pm 0.5$~nm and $37$ $\pm 1$ \%, respectively.

In the bulk state 7CB exhibits an isotropic-to-nematic ordering transition at a temperature $T_{\rm IN}$=315~K. Given the low vapor pressure of 7CB at room temperature and the channel geometry with large, inner surfaces we cannot fill the matrix via the gas phase. Therefore, we imbibe it with eight binary 7CB/acetone solutions \cite{Gruener2011}. After evaporation of the high vapor pressure solvent, the matrix is filled with the remaining, low-vapor pressure liquid crystals up to a filling fraction $f$, that depends on the initial solute concentration. \textcolor{black}{The filling fraction after evaporation is verified by weighting of the partially filled matrix and a comparison with the mass of the empty matrix of 20 mg. By a gradual increase of the solute concentration we prepare 10 filling fractions in the range from $f=$0.1 to 1 at room temperature. The evaporation of the solvent is controlled by time-dependent mass measurements of the sample at a temperature of 325 K and followed an exponential decay with time.  After 1 hour, the acetone had evaporated to 1/100 of its initial concentration, and thus for the low-$f$ samples the further mass drop was below the detection limits of our balance, which was $5 \cdot 10^{-5}$ g. Then we waited for another approx. 3 hours before we started the birefringence measurements in order to guarantee the absence of any significant amount of solvent in the sample. The estimated error margin in $f$ due to the uncertainties in the mass determination amounted to 0.005.}

\begin{figure*}[htbp]
\epsfig{file=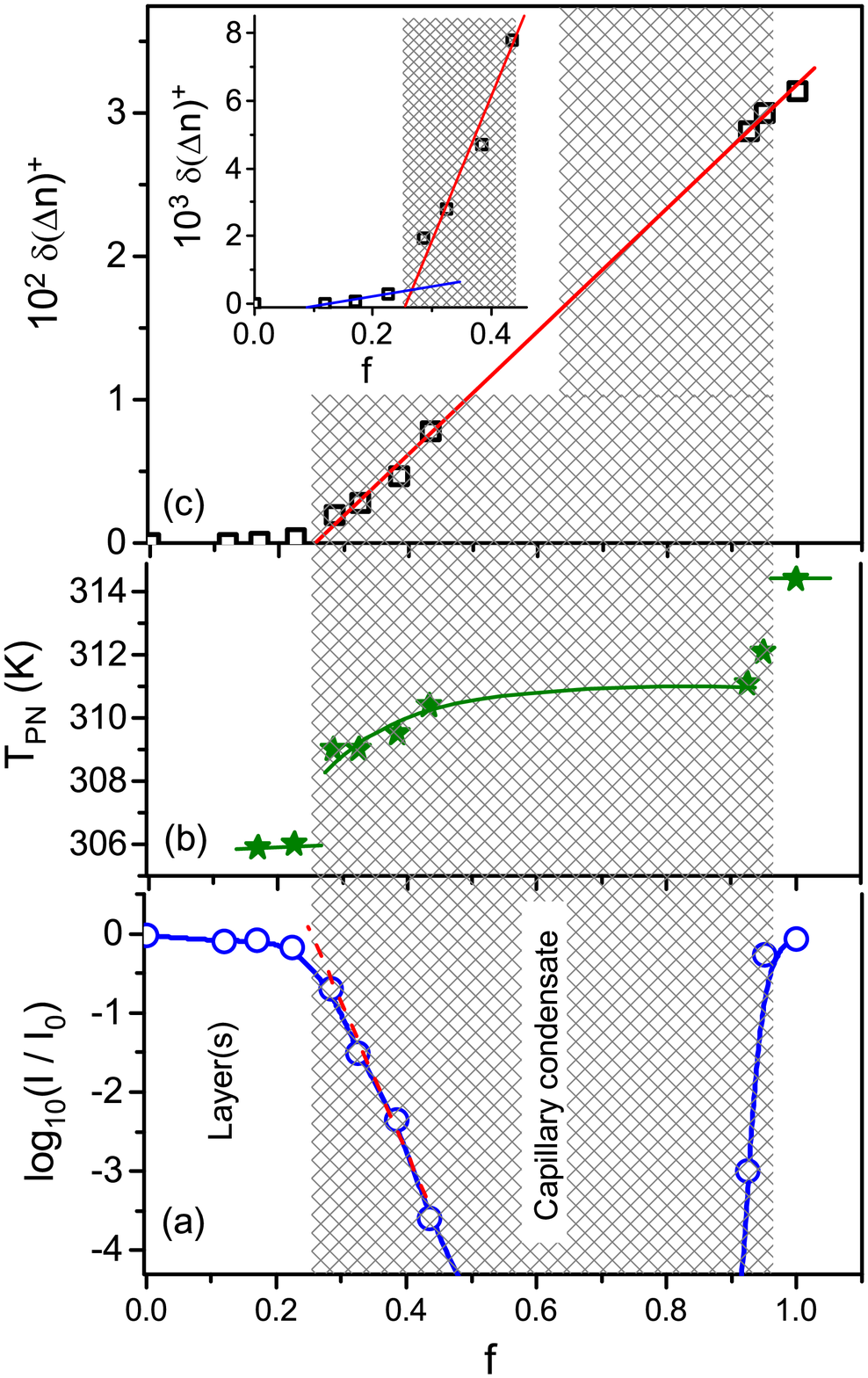, angle=0, width=0.9\columnwidth}
\epsfig{file=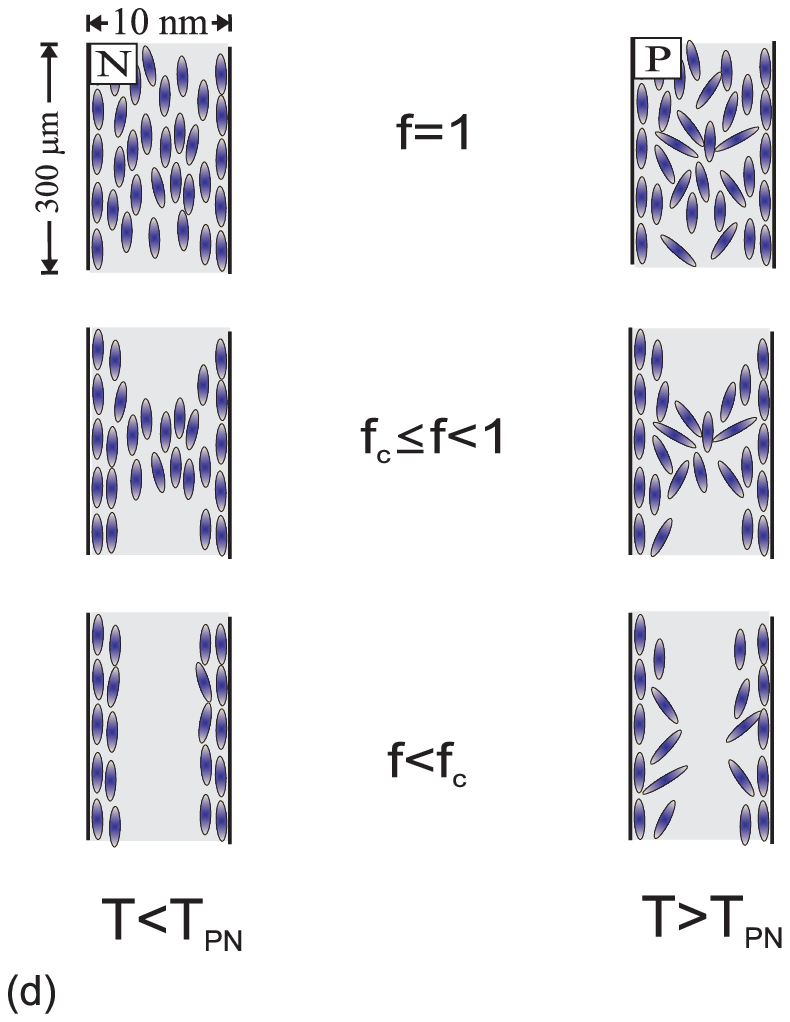, angle=0, width=1\columnwidth}
\caption{(Color online). 7CB in nanoporous silica. (a) Light transmission $log_{\rm 10}(I/I_{\rm 0})$ vs. filling fraction $f$, (b) Temperature of the paranematic-to-nematic transition $T_{\rm PN}$ vs. filling fraction $f$ as determined from the results of optical birefringence measurements shown in Fig.~\ref{fig2}, (c) $f$-dependence of the effective optical birefringence $\delta (\Delta n)^+$ at $T=T_{\rm PN}-10$~K as discussed in the text, (d) schematic side views of three characteristic capillary filling regimes: adsorbed monolayer(s), film regime ($f<f_{\rm c}$), capillary condensate ($f_{\rm c}\le f<1$) and completely filled substrate ($f=1$). The molecular orientations in the different filling regimes are illustrated below (left panel d) and above (right panel d) $T_{\rm PN}$, that is in the nematic and paranematic state of the confined liquid crystal. In the panels (a-c) the $f$-range typical of capillary condensation is shaded. The solid lines in panel (a) and (b) are guide for the eyes, whereas in panel (c) linear fits are presented.} \label{fig1}

\end{figure*}

\section{Results and discussion}
It is instructive to compare this filling procedure with the standard way for the preparation of partial liquid fillings of nanoporous substrates via the gas phase, i.e. vapor sorption isotherms. For nanoporous solids they exhibit distinct regimes typical of characteristic arrangements of the pore liquid: At small fractional filling $f$, below a certain threshold value $f_{\rm c}$, an adsorbed film is homogeneously formed at the pore walls, see sketch in Fig.~\ref{fig1}d, $f<f_{\rm c}$. For $f\ge f_{\rm c}$ the pores fill via capillary condensation, that is the formation of liquid bridges terminated by concave menisci. Both regimes can be inferred by light adsorption experiments. Upon onset of capillary condensation or evaporation typically a drastic decrease in optical transmission, $\lg(I/I_{\rm 0})$ is observable \cite{Page1993, Kityk2009, Soprunyuk2003} (Here $I_{\rm 0}$ and $I$ are the intensities of incident and transmitted laser light, respectively). The matrix turns milky and the reduction in the transmitted light originates in scattering at the capillary bridges/void structures, which have sizes on the length scale of the wavelength of visible light \cite{Naumov2008, Naumov2009}. By contrast, the matrix remains transparent both in the film condensed state ($f<f_{\rm c}$) and in the completely filled state ($f\sim$1), where such structures are absent.

Laser light adsorption experiments on our solute-prepared samples exhibit an analogous behavior as a function of $f$,  see Fig.~\ref{fig1}a. A sizeable decrease in light transmission is observed for $f \ge f_{\rm c}=0.25$. The gradual disappearance of the bubbles and/or capillary bridges upon reaching a complete filling is indicated for $f > 0.95-0.97$ and, eventually the full optical transmission upon vanishing of all empty pore segments is recovered for $f \rightarrow 1$. Note that the transmission drops by four orders of magnitude in the capillary condensate regime, eventually even below the sensitivity of our setup.


In contrast to nanoporous silica powders or disordered, sponge-like pore structures, our monolithic silica membrane is particularly suitable in order to evaluate the collective orientational order of liquid crystals by optical birefringence measurements in a transmission geometry \cite{Kityk2008}. In the bulk nematic state the optical birefringence changes linearly with the orientational order parameter, $Q=\frac{1}{2}\langle 3 \cos^2\Theta-1\rangle$, where $\Theta$ is the angle between the long axis of a single molecule and a direction of preferred orientation of that axis, the director \cite{Kumar2001}. The brackets denote an averaging over all molecules under consideration.

The simple relation between optical birefringence and $Q$ the bulk system is also conserved for the liquid crystal/matrix composite material, provided the geometrical birefringence is subtracted \cite{Kityk2009}.  \textcolor{black}{Due to the changes in the geometrical arrangement of the liquid upon gradual filling, the latter one changes considerably in the entire region of partial fillings. This holds even in the case of the adsorption of a liquid with isotropic building blocks, like neopentane, which we experimentally and theoretically demonstrated in two previous studies \cite{Kityk2009, Wolff2010}. It is, however, practically $T$-independent for a fixed $f$. For this reason the change of the optical birefringence, $\delta (\Delta n)$ versus $T$, represents the quantity which most properly characterizes the orientational ordering inside the porous substrate.} It was measured with a self-designed, high-resolution polarimetry setup presented in Refs. \cite{Skarabot1998, Kityk2008}. The high sensitivity of this device, which employs a dual lock-in detection scheme, allows one to obtain data even for low optical transmissions of the matrix. Therefore, it is possible to gain reliable data even upon onset of capillary condensation and upon reaching complete filling, as long as the transmission is larger than 10$^{-4}$.

\begin{figure}[htbp]
\epsfig{file=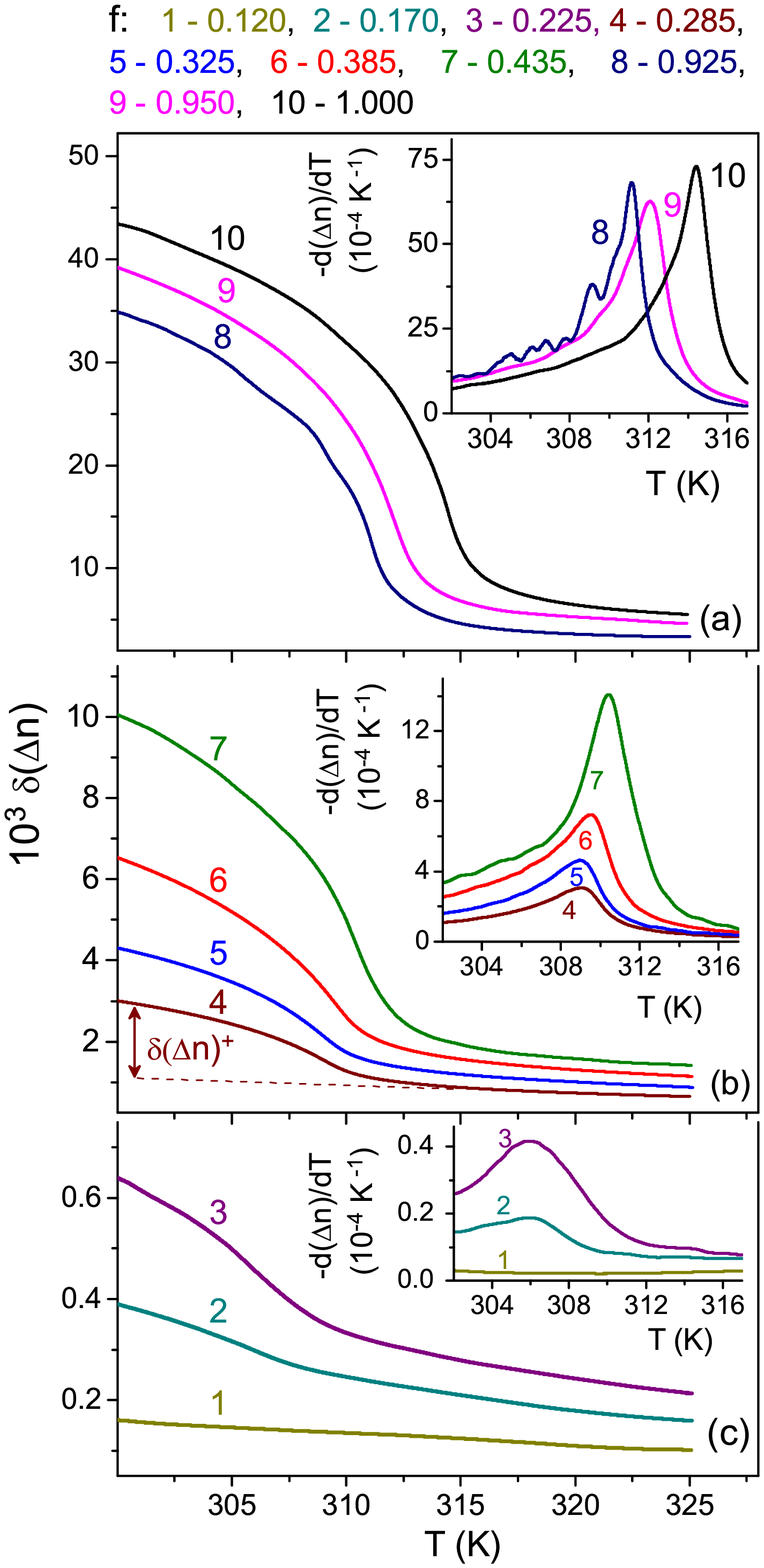, angle=0,width=0.9\columnwidth}
 \caption{(Color online). Temperature-dependent changes of the optical birefringence, $\delta (\Delta n)$, and its temperature derivatives (insets), $-d(\Delta n)/dT$, for selected filling fractions $f$ as referenced in the header of the figure. Panels (a) and (b) correspond to the capillary-condensed regime and panel (c) represents the the film-condensed regime. Note that the waviness  of $\delta (\Delta n)$ for f=0.925 results from the significant drop in transmitted light for that filling fraction - see Fig. 1. \textcolor{black}{The dashed line in panel (b) examplifies a linear extrapolation of $\delta (\Delta n)$ of the paranematic state, which is done for each filling fraction $f$ in order to determine $\delta (\Delta n)^+$.}}
\label{fig2}
\end{figure}

In Fig.~\ref{fig2}c $\delta(\Delta n)$ is plotted versus temperature for a selected set of $f$'s in the film-condensed state. There exists a final birefringence and thus orientational order at high-$T$ in all measurements. It is typical of a sizable collective ordering of the molecules induced by the interaction with the wall (interface anchoring). Silica surfaces favor a planar anchoring in the case of the family of cyanobiphenyl liquid crystals (CBs), as documented in experiments \cite{Vandenbrouck1998, DrevensekOlenik2003} and atomistic computer simulations \cite{Pizzirusso2012}. The magnitude and the positive sign of the changes of the optical birefringence for the lowest filling fraction ($f=$0.12, corresponding to a partially filled, first adsorbed monolayer) indicates indeed such an alignment. It is even compatible with a preferred orientation of the long axis of the liquid crystals along the axial direction of the nano channels - as illustrated in Fig.~\ref{fig1}d. Note that there is a small, gradual increase $\delta(\Delta n)(T)$ upon cooling for this filling fraction, but no indication of a distinct change in collective orientational order. 

If one assumes a perfect cylindrical geometry and an intermolecular distance of 0.4~nm \cite{Komolkin1994}, the molecules added in the next filling steps correspond to $\sim$1.5  and 2 monolayers  ($f=$0.17 and 0.225). They exhibit sizeable increases in $\delta(\Delta n)(T)$ upon cooling, which are characteristic of a rearrangement between partial collective orientational order (a paranematic state) to full collective orientational order (nematic order). The symmetry of those states is identical; the transition is no ''phase transition'' in a classical sense. It is, however, possible to define $T_{\rm PN}$  as a temperature of most pronounced changes of $Q(T)$, in analogy to the ''true" phase transition temperature. We determine this quantity by a calculation of the maximum in the $T$-derivative of $\delta(\Delta n)(T)$ - see insets in Fig.~\ref{fig2}. For the nanometric liquid crystal film it yields a $T_{\rm PN}$ of 306~K. This downward-shift $\Delta T_{\rm IN}$ of 10K in respect to the bulk system compares surprisingly well with an expected $\Delta T_{\rm IN}$ of $\sim$ 9~K, which one achieves by an extrapolation of thickness-dependent boundary effects, explored in thin-film geometry on silica (with film thicknesses of several hundred nanometers) \cite{Yokoyama1988}, down to the nano metric films investigated here. The gradual increase in thermotropically induced molecular reorientations upon film-thickness increase, observed here, corroborates a continuous increase in molecular orientational freedom with increasing distance from the pore walls documented in computer simulations for cylindrical nanopores \cite{Ji2009a, Ji2009b}.

\textcolor{black}{The evolution of the thermotropic changes upon formation of capillary bridges ($f_{\rm c} \ge f<1 $) is documented in the panels (a) and (b) of Fig.~\ref{fig2}. The  transition from paranematic-to-nematic state remains continuous in this range of filling fractions whereas the temperature behavior of the optical birefringence and its relevant changes systematically scale with $f$. In order to characterize these changes quantitatively we introduce the quantity, $\delta (\Delta n)^+$ as the difference between the measured $\delta (\Delta n)$ at the lowest temperature investigated and its linearly extrapolated value from the paranematic phase far above $T_{PN}$, see dashed line in Fig. \ref{fig2}b. By this subtraction we remove any contribution purely induced by the changes in the geometric birefringence upon capillary filling. Therefore, this quantity represents an effective order parameter $\bar{Q}=V^{-1}\int_V{Q(r)dV}$, characteristic of the integrated strength of the thermotropic, collective orientational (nematic) ordering inside the channels.}


In Fig.~\ref{fig1}c the $f$-dependence of $\delta (\Delta n)^+$ is shown as determined in the confined nematic phase at $T=T_{\rm IN}-10$K for each selected $f$. It increases linearly with $f$ both in the film-condensed and in the capillary-condensed state. It is striking, however, that the slope in the experimentally accessible capillary-condensed regime (0.25$\leq$f$\leq$0.98) is by more than one order of magnitude larger than in the film-condensed state - see also inset in Fig.~\ref{fig1}c. This means that the film-condensed and capillary-condensed state is clearly distinguishable in terms of the evolution of thermotropic orientational order: In agreement with molecular simulation studies \cite{Ji2009, Ji2009b}, the freedom in orientational molecular ordering is considerably larger near the pore axis (i.e. in the core of the pore filling) than in the proximity of the channel wall, where the molecules are much more strongly influenced both by channel-wall roughness (quenched disorder) and the anchoring field of the channel walls  \cite{Ji2009, Ji2009b, Pizzirusso2012}.

The linear increase of $\delta (\Delta n)^+$ with $f$ in the film-condensed and capillary-condensed regime suggests that material added in one step does not affect the reorientational behavior of the material adsorbed in the previous filling steps. We can not exclude, however, that the formation of capillary condensate in the pore center and the corresponding vanishing cylindrical liquid-vapor interfaces affects the reorientation behavior of the molecules in the film-condensed, boundary layer state.

The temperature of the paranematic-to-nematic transition exhibits only a weak increase in the entire region of capillary condensation ($f_{\rm c}\leq f <  1$) and remains at least $\sim$ 4~K below the transition point $T_{\rm PN}$ of the fully filled ($f=1$) porous substrate, see Fig.~\ref{fig1}b. 

Note, however, that there is a sizable upward jump of $T_{\rm PN}$ between the slightly underfilled ($f<1$) and the completely filled state ($f=1$) - see Fig.~\ref{fig1}b. Hence, $T_{\rm PN}$ of the continuous transition of the completely filled state is close to the temperature of the discontinuous bulk transition $T_{\rm IN}$, in agreement with our previous experimental study and phenomenological treatment of this state with a Landau-de-Gennes model enriched by confinement and quenched disorder effects (KKLZ-model)  \cite{Kutnjak2003, Kutnjak2004, Kityk2008}.

The abrupt $T$-shift in the transition temperature may result from the complete vanishing of any liquid/vapor interface and the resulting reduction of inhomogeneities in the system, which rather favor the disordered paranematic state. \textcolor{black}{In particular, for free liquid/vapor interfaces, which are present for $f<1$, the rod-like molecules are aligned parallel to the surface normal and the PN-transition in their vicinity is rather shifted upwards in $T$  \cite{Kasten1995}. Albeit because of the high curvature of the menisci in the pores this anchoring condition competes with elastic energies, which effectively may lead to a reduction of $T_{\rm PN}$. Without having a detailed microscopic information of the orientational structure, a quantitative estimation of these influences appears difficult, in particular since at the triple line solid/vapor/liquid crystal of the menisci perimeters peculiar defect structures may occur. Molecular dynamics simulations on confined rod-like crystalline systems could be helpful to gain quantitative insights for this complex problem.} 

There is another alternative explanation, which is related to the hydrostatics of the confined liquid. In the capillary-condensed state the liquid experiences a tensile pressure, dictated by the concave curvature of the menisci terminating the liquid bridges. It is known that this negative hydrostatic pressure causes not only subtle deformations of the rigid, nanoporous matrix \cite{Guenther2008, Prass2009, Gor2010}, it also significantly affects density, and thus pressure-dependent first-order transitions, most prominently the liquid-solid transition \cite{Morishige2006, Schaefer2008, Moerz2012}. Therefore, it is in principal also expected to affect the isotropic-to-nematic transition \cite{Rein1993, Manjuladevi2002}. According to the Laplace formula applied to the menisci in the capillaries, while assuming good wetting conditions (mean curvature radius = - pore radius = -6 nm, surface tension of 7 CB=31.7~mN/m \cite{Delabre2009}) the tensile pressure in the liquid bridges amounts to $\sim$-10.5 MPa for $f_{\rm c}<f<1$. This pressure is completely released upon reaching $f=1$ (mean curvature radius = $\infty$). An extrapolation of the pressure-dependence of $T_{\rm IN}$ of 7CB, reported in the literature \cite{Rein1993},  towards negative pressures yields a corresponding downward shift of 3.8~K between partially and completely filled state, in good quantitative agreement with the observed $T$-shift. It is interesting to note that this $T$-shift is much larger than any tensile pressure effect on liquid crystals that has been reported to date. Presumably, this results from the extreme challenges \cite{Manjuladevi2002} associated with experiments aimed at reaching tensile pressures comparable to the ones established here in the partially filled state. Since the true curvature of the menisci at high $f$-values is not accessible in our experiments and since studies with higher f-resolution are experimentally demanding both due to the sizable light scattering in that regime and because of the error bars in the filling fraction (resulting from $T$-dependent volume changes in the liquid), this conclusion remains somewhat speculative and more rigorous studies are planned for the future.

\section{Summary}

In summary our optical studies allowed us to experimentally document an inhomogeneous thermotropic orientational order behavior upon sequential filling of silica nano-capillaries with a rod-like liquid crystal which encompasses four distinct regimes: the monolayer state, the film-condensed state, the capillary-condensed regime and the complete filling.

Since orientational ordering is key to functional materials with switching capability, we believe that our observations are not only of high fundamental interest, but also of importance for the emerging field of nanophotonics \cite{Abdulhalim2012}: Our study documents that the thermooptical properties of nanoporous systems (or single nano-capillaries) can be tailored over a surprisingly wide range simply by variation of the the filling fraction with liquid crystals.
 
For the future we envision analogous experimental studies on the molecular dynamics, which should show a similar complex behavior as the static properties investigated here. We also hope that our findings stimulate temperature-dependent simulation studies and model calculations on partially filled nano-capillaries, aimed at an exploration of the remarkably complex phase transition phenomenologies revealed here as a function of material distribution in the capillaries. 

\section{Acknowledgment}
This work has been supported by the Polish National Science Centre under the Project "Molecular structure and dynamics of liquid crystals
based nanocomposites" (decision No DEC-2012/05/B/ST3/02782) and by the German Research Foundation (DFG) by Grant No. Hu850/3-1.

\bibliographystyle{apsrev}

\end{document}